\documentclass[12pt,a4paper,english]{article}
\usepackage[T1]{fontenc}
\usepackage[latin1]{inputenc}
\usepackage{float}
\usepackage{graphicx}

\makeatletter

\usepackage{graphicx}

\usepackage{babel}
\makeatother
\begin{document}
\begin{center}November, 2005\\
 \begin{large}
{\textbf{Crystal-field splittings in  CeX (X= N, P, As, Sb, Bi) compounds}}\\
\vspace{.4cm}
P. Roura-Bas~\(^{a,*}\) ,V. Vildosola~ \(^{a,b}\)~~and A. M. Llois~\(^{a,b}\)\\
\end{large} \vspace{0.2cm} \begin{small}
{\textit{\(^a\) Departamento de F\'{i}sica, Centro At\'{o}mico Constituyentes, Comisi\'{o}n 
Nacional de Energ\'{i}a At\'{o}mica}}\\
{\textit{\(^b\) Departamento de F\'{i}sica, Universidad de Buenos Aires, Buenos Aires, Argentina}}\\
\end{small}\end{center}

\vspace{1cm}

\begin{small}

The unusual and interesting physical properties of rare earth intemetallic compounds have their origin in the combination of strongly correlated 4f states and their hybridization with the conduction electron sea, which gives rise to their complex low temperature Kondo behavior. In particular, Ce compounds are very sensitive to the crystalline and chemical environment, as compared to other rare earth systems. The interaction of the 4f state with the conduction band plays an important role in the determination of the different magnetic, structural and transport properties of these systems. Among the cerium compounds, those of the type CeX, which crystallize in the rock salt structure, exhibit extremely unusual magnetic properties. By making use of the mixed LDA-NCA calculation technique we analyse the crystal-field splittings of CeX compounds (X=N, P, As, Sb, Bi). The obtained ab-initio hybridization functions are taken as imputs to calculate the crystal-field splittings within NCA (non crossing approximation) and the tendencies are contrasted with experiments.
\vspace{0.3cm}
\(^{*}\) Corresponding author\\
CAC-CNEA, Av. Gral. Paz 1499, San Mart\'in (1650), Buenos Aires, Argentina\\
roura@tandar.cnea.gov.ar
\end{small}

\vspace{1cm}

Cerium monopnictides exhibit anomalous magnetic properties due
to the magnetic anisotropy and to the particular hybridization strength of
the 4f electron with the conduction band. The crystal-fields splittings
($\Delta_{CF}$) of the 4f level in the heavier monopnictides (CeBi and CeSb)
are considerably smaller than in the lighter Ce monopnictides (CeAs,
CeP). This behaviour was previously studied and understood by Wills
and Cooper \cite{Wills-cooper-1}. The dominant contribution to the
splittings in these monopnictides can be obtained from the point-charge
(PC) model which is appropiate for insulators or ionic systems. This PC model alone can
account for the CF splittings of Rare Earth monopnictides when the
Rare Earth goes from Pr to Tb but it fails to describe Ce monopnictides
because in Ce compounds the 4f-band hybridization cannot be neglected.\\
Wills and Cooper consider that the total $\Delta_{CF}$  in these CeX
systems is the result of two independent contributions with different
sign: the extrapolated value from the PC model with non-hybridized
4f levels and the splittings induced by hybridization. The former gives 
rise to positive splittings which increase from
Bi to P. In the cubic point group, the multiplet $J=\frac{5}{2}$
decomposes into the $\mid\frac{5}{2};\Gamma_{7}\rangle$ doublet and
the $\mid\frac{5}{2};\Gamma_{8}\rangle$ quartet. So that a positive
 splitting means that $\Gamma_{7}$ is the ground state. They calculate 
the 4f hybridization function out of the conduction
band density of states which is obtained from first principles, using
the linear muffin-tin-orbital (LMTO) method and the self-consistent
potential within the atomic-sphere approximation (ASA) which does
not contemplate the anisotropy of the crystalline environment. Within
these calculations the 4f state is treated as a core state. Then,
they calculate the hybridization contribution to $\Delta_{CF}$ 
for CeX (X=P, As, Sb and Bi) on the basis of the Anderson Hamiltonian
using second order pertubation theory. These hybridization contributions
have an opposite sign to that of the PC model. In this way, they can explain 
the suppresion of the crystal-field splitting of the heavier Ce monopnictides observed
from the experimental data. 

In the present work, we calculate $\Delta_{CF}$ for this series of monopnictides
in a different way. We perform \textit{ab initio} calculations using
the Full Potential-LAPW method within the local density approximation
(LDA) \cite{wien} instead of the ASA potential and we treat the 4f states
as part of the valence band. We then calculate the projected $4f$-densities
of states, $\rho_{mm'}$, for the systems at the experimental volumes
(where $m$ runs for the seven irreducible representations for a cubic environment)
and compute the hybridization function, $\Gamma_{mm'}(\epsilon)$,
following ref. \cite{Gunnarsson,Han}. 

The different $\Gamma_{mm'}(\epsilon)$ contain detailed information
of the electronic structure of each system and are used as input for
the Anderson Impurity model which is solved within an extended non-crossing
approximation (NCA). In the Anderson model the value of the \textit{bare}
$f$ energy level, $\varepsilon_{f}$, is taken as the experimental
value that comes out of photoemission spectra \cite{photoem}. This LDA-NCA
approach has already been applied to cubic and tetragonal $Ce$ based
systems to calculate $\Delta_{CF}$'s and the trend in the Kondo energy scales yelding results 
in good agreement with the available experimental information \cite{Han,Vildosola}. 

Within this technique  $\Delta_{CF}$'s arise from band-f
hybridization and the anisotropy of the crystalline environment. 

Experimental data of the CF splittings for the compounds under study
in the present work are the following: 8 K for $CeBi$, 37 K for $CeSb$,
159 K for $CeAs$ and 172 K for $CeP$. All the compounds have been
reported to have the $\Gamma_{7}$ multiplet as the ground state \cite{7}. 
There is no available data for $CeN$. 

Figure 1 and 2 show the calculated hybridization function for $CeP$
and $CeSb$. In the inset, the detailed structure of these functions
around the Fermi energy is shown and the $\Gamma_{7}$ and $\Gamma_{8}$ average values are explicitly shown. It can be observed that while the
unoccupied states are governed by the $\Gamma_{8}$ symmetry, close
to the Fermi level the $\Gamma_{7}$ one is stronger. As it is
discussed in another contribution to this conference \cite{roura-cey}, the 5d-Ce band hybridizes mainly with the $\Gamma_{7}$ 4f
states while the p-X band with the $\Gamma_{8}$ 4f states.\\
Whithin this model, a stronger $\Gamma_{8}$ hybridization
would imply a $\Gamma_{8}$ symmetry for the ground state. If we consider
the CF splittings that result from the LDA calculations taking into
account the whole energy range for the conduction band (up to 1 Ry approx.), 
we obtain that the ground state is $\Gamma_{8}$ for
all the CeX series and a decreasing  $\Delta_{CF}$ from
N to Bi, the value of the splittings being much larger than the experimental data. 
The reason why these results disagree both in sign and magnitud of  $\Delta_{CF}$
is that the calculated hybridization function for
the p-X band ($\Gamma_{8}$) is overestimated due to the fact that
the LDA calculation does not account for the strong ionic character
of these monopnictides compounds. Wills and Cooper also obtain a strong
hybridization for the p-X band which is compensated by taking into account
the CF shift from the ionic point charge model. In the present work
we are able to reproduce the correct trend of  $\Delta_{CF}$'s
if we consider the energy spectrum of the bands up to an energy of the
order of 6000 K above the Fermi level, discarding the unoccupied states
above that energy. With this criterion, we obtain a $\Gamma_{7}$
ground state for all the series and the calculated values for the
CF splittings are: 160K for CeP, 150K for CeAs, 70K
for CeSb and 50 K for CeBi. For CeN we also obtain a $\Gamma_{7}$ ground state and a splitting greater that room temperture. 

As it is discussed in Ref. \cite{roura-cey} the more important contribution to the  $\Delta_{CF}$'s in this series of compounds 
comes from the energy dependent hybridization function around the Fermi level ($E_F$). 
Within this energy range, $\Gamma_{7}$ hybridization is stronger and
the difference between the $\Gamma_{7}$ and $\Gamma_{8}$ hybridization
strength varies along the series because within the Full Potential-LDA
calculation we take the 4f states as part of the valence band giving
rise to different CF splittings. We claim that the main contribution
to the hybridization around $E_F$ comes from the 5d-4f interaction. In the systems
where the 4f states are more localized (CeBi, CeSb) the 5d-4f hybridization
is smaller and in turn, a smaller CF splitting is obtained. We consider
that the p-4f hybridization should not be that strong for these ionic compounds. 

In previous calculations \cite{Han,Vildosola}, it was not necessary
to cut the energy spectrum because those compounds (the cubic $CeM_3$ and 
the tetragonal $CeM_2Si_2$) do not present
such a strong ionic character and LDA can give an appropiate description
of the conduction bands.

This work was partially funded by UBACyT-X115, PICT-0310698 and PIP 2005-2006 Num. 6016. A. M. Llois and V. L. Vildosola belong to CONICET (Argentina).

\begin{figure}[H]
\begin{center}\includegraphics[%
  scale=0.5,
  angle=270]{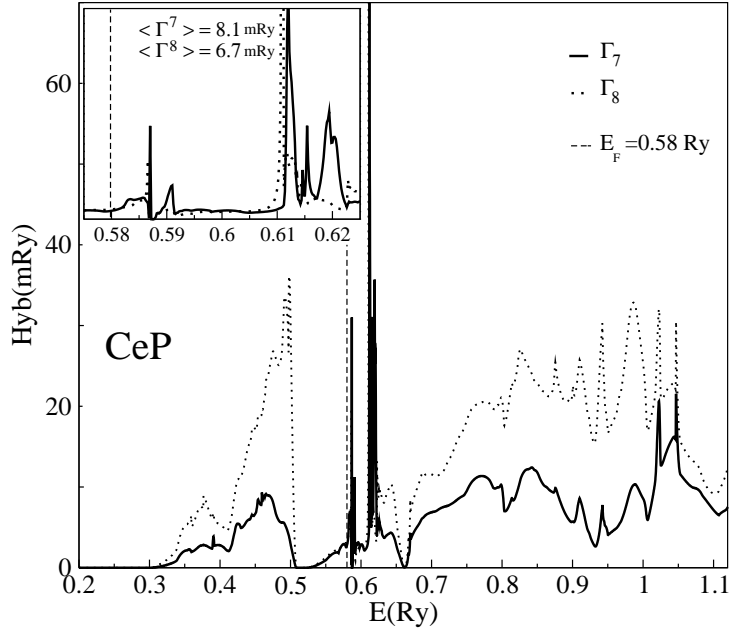}\end{center}

\caption{The calculated hybridization function for CeP between the conduction band and the $\Gamma_7$ (solid curve) and $\Gamma_8$ (dashed curve) 4f states. In the inset  a detail of the hybridization around the Fermi level.}
\end{figure}

\begin{figure}[H]
\begin{center}\includegraphics[%
  scale=0.5,
  angle=270]{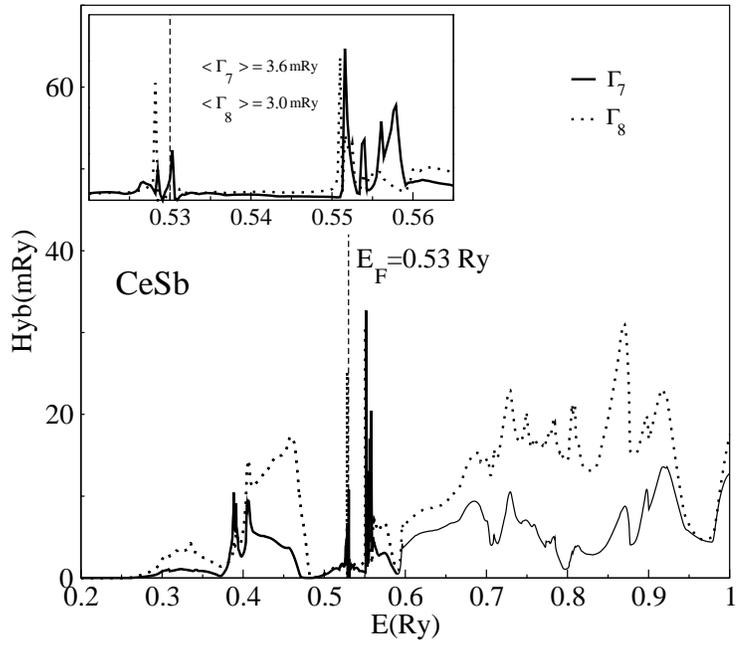}\end{center}

\caption{{\it Idem} than in Fig. 1 but for CeSb.}
\end{figure}

\end{document}